\documentclass[prl,aps,twocolumn,superscriptaddress,showpacs,preprintnumbers]{revtex4-1}
\usepackage{amsmath,amssymb}
\usepackage{graphicx}
\usepackage{tabularx}
\newcolumntype{Y}{>{\centering\arraybackslash}X}

\begin{document} 

\preprint{EFI 16-6}
\title{Higher-Spin Theory of the Magnetorotons} 

\author{Siavash Golkar}
\affiliation{Department of Applied Mathematics and Theoretical Physics, University of Cambridge, Cambridge CB3 0WA, United Kingdom}
\author{Dung Xuan Nguyen}
\affiliation{Department of Physics, University of Chicago, Chicago, Illinois 60637, USA}
\author{Matthew M.~Roberts}
\affiliation{Kadanoff Center for Theoretical Physics, University of Chicago, Chicago, Illinois 60637, USA}
\author{Dam Thanh Son}
\affiliation{Department of Physics, University of Chicago, Chicago, Illinois 60637, USA}
\affiliation{Kadanoff Center for Theoretical Physics, University of Chicago, Chicago, Illinois 60637, USA}

\date{February 2016}

\begin{abstract}
Fractional quantum Hall liquids exhibit a rich set of excitations, the
lowest-energy of which are the magnetorotons with dispersion minima
at a finite momentum.
%Recent works have established nontrivial
%connection of the half filled Landau level with spin liquids and
%interacting topological insulators.
We propose a theory of the
magnetorotons on the quantum Hall plateaux near half filling, namely,
at filling fractions $\nu=N/(2N+1)$ at large $N$.  The theory involves
an infinite number of bosonic fields arising from bosonizing the
fluctuations of the shape of the composite Fermi surface.  At zero
momentum there are $O(N)$ neutral excitations, each carrying a
well-defined spin that runs integer values $2,3,\ldots$.  The mixing
of modes at nonzero momentum $q$ leads to the characteristic bending down
of the lowest excitation and the appearance of the magnetoroton
minima.  A purely algebraic argument shows that the magnetoroton
minima are located at $q\ell_B=z_i/(2N+1)$, where $\ell_B$ is the
magnetic length and $z_i$ are the zeros of the Bessel function $J_1$,
independent of the microscopic details. We argue that these minima are
universal features of any two-dimensional Fermi surface coupled
to a gauge field in a small background magnetic field.
\end{abstract}
\pacs{73.43.Cd,73.43.Lp}

\maketitle 

Interacting electrons moving in two dimensions
in a strong magnetic field can form nontrivial topological states: the
fractional quantum Hall liquids~\cite{Tsui:1982yy,Laughlin:1983fy}.
When the lowest Landau level is filled at certain rational filling
fractions, including $\nu=N/(2N+1)$ and $\nu=(N+1)/(2N+1)$ (Jain's
sequences), the quantum Hall liquid is gapped, and the lowest-energy
mode is a neutral mode.  Girvin, MacDonald, and
Platzman~\cite{Girvin:1986zz} proposed, based on a variational ansatz,
that the neutral excitation has a broad minimum at $q\ell_B\sim 1$ at
the Laughlin plateau $\nu=1/3$.  Several years later, the existence of
a neutral mode was confirmed experimentally~\cite{Pinczuk:1993}.
Later experiments revealed a surprising richness in the structure of the
spectrum of neutral excitations.  Unexpectedly, the $\nu=1/3$ state
may have more than one branch of
excitations~\cite{Hirjibehedin:2005}.  Furthermore, higher in the Jain
sequence, i.e., for $\nu=2/5$, $3/7$, etc., the lowest excitation has
been found to have a dispersion with more than one
minima~\cite{Kang:2000,Kukushkin:2009}.  Various theoretical
approaches have been brought to the problem of the
magnetoroton~\cite{Zhang:1992,Simon:1993,Scarola:2000,Majumder:2009,Yang:2012}.
Currently, the most common viewpoint is based on the composite fermion
picture of the fractional quantum Hall effect, in which the neutral
modes are bound states of a composite fermion and a composite hole.

The notion of the composite fermion is tightly connected to the
Halperin-Lee-Read (HLR) field theory~\cite{Halperin:1992mh}, proposed
as the low-energy description of the half filled Landau level.
Recently, an analysis of the particle-hole symmetry of the lowest
Landau level has lead to a revision of the HLR proposal: the
low-energy degrees of freedom is now a Dirac composite fermion coupled
to a gauge field~\cite{Son:2015xqa}.  Magnetorotons provide a rare
window into the dynamics of a Fermi surface coupled to a gauge field,
a long-standing problem of condensed matter
physics~\cite{Altshuler:1994zz,Polchinski:1993ii}.

None of the previous analytical approaches to the magnetoroton can
deal with the non-Fermi liquid at $\nu=1/2$, or even with a composite
Fermi liquid with general nonzero values of the Landau parameters. In this
Letter, we develop a theory of neutral excitations in the quantum Hall
liquid, reliable in the limit $N\to\infty$ in Jain's series
$\nu=N/(2N+1)$, where quantum Hall plateaux
have been found to up to at least $N=10$~\cite{Pan:2008}.  In this
theory, the neutral excitations are viewed as quantized shape
fluctuations of the Fermi surface.  This interpretation is quite
different from what has been suggested so far and is one
with a predictive power.  In particular, one can relate the whole dispersion
curves of the neutral excitations to the excitation energies at zero
momentum.  We find that the dispersion curves have deep magnetoroton
minima at large $N$.  Remarkably, the momenta at the magnetoroton
minima are independent of all microscopic dynamics and are in
quantitative agreement with existing experimental data even for small
$N$.

{\em Quantizing the shape of the Fermi surface.}---To find the
magnetorotons we will first bosonize the Fermi surface.  This
procedure was studied
previously~\cite{Haldane:1994,Houghton:1992dz,CastroNeto:1994,Mross:2011}.
%but merits a reconsideration due to the existence of the background
%magnetic field and the gauge interactions in our composite fermion
%system.
Our approach relies on a commutation algebra of fluctuations of the shape
of the Fermi surface, first derived by Haldane~\cite{Haldane:1994}.  Here
we provide a simple semiclassical derivation of this algebra.

We assume that the $\nu=1/2$ state is gapless and has a Fermi surface
with the Fermi momentum $p_F$, related to the external magnetic field
$B$ by $p_F^2=B$.  The Fermi liquid is characterized by the Fermi
velocity $v_F$ and Landau's parameters $F_n$.  The effective mass
is defined as $m_*=p_F/v_F$, the Fermi energy scale as
$\epsilon_F=v_Fp_F$.

\begin{figure}
  \centering
  \includegraphics[width=16em]{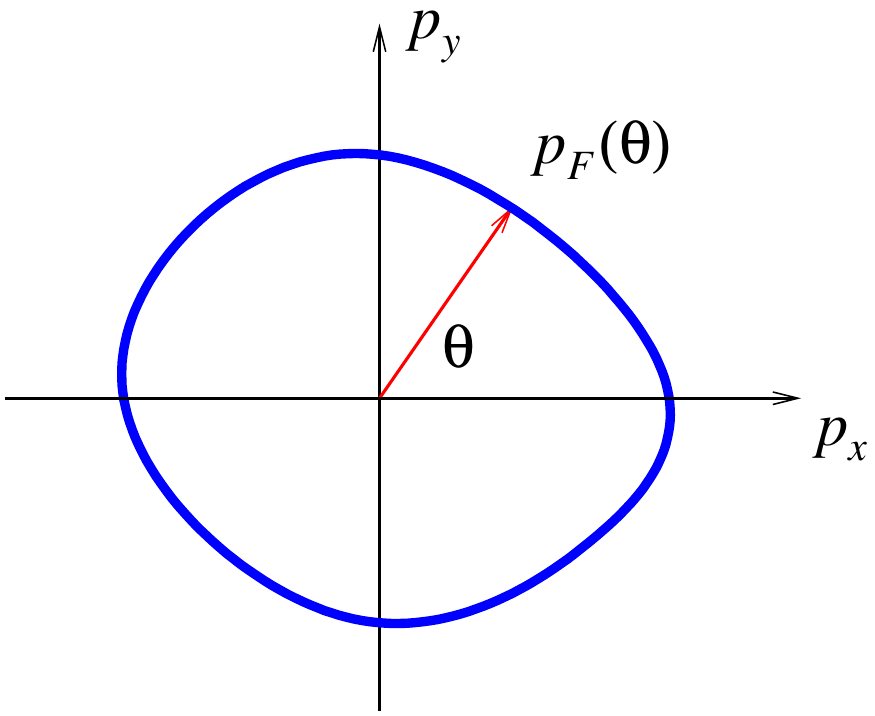}
  \caption{A deformed Fermi surface.}
  \label{fig:fs}
\end{figure}

In the fractional quantum Hall $\nu=N/(2N+1)$ state, the composite
fermions live in a magnetic field $b=B/(2N+1)$, effectively forming an
integer quantum Hall state.  We are interested in the regime of
frequency and momentum of the order of $N^{-1}$ compared to the Fermi energy
and momentum.  We now propose that all low-energy excitations can be
viewed as deformations of the Fermi surface from the circular shape,
which we parameterize by a function $p_F(t,{\bf x},\theta)$ that
depends on time and space and also on the direction in momentum space
$\theta$ ($p_y/p_x=\tan\theta$) (see Fig.~\ref{fig:fs}).  Furthermore,
we decompose the perturbation into different angular momentum
channels:
\begin{equation}
  p_F(t,{\bf x},\theta) = p_F^0+u(t,{\bf x},\theta)
  = p_F^0 + \!\!\! \sum_{n=-\infty}^\infty\!
  u_n(t,{\bf x})\, e^{-in\theta}.
\end{equation}
In the language of Landau's Fermi liquid theory, the state
parameterized by $p_F(t,{\bf x},\theta)$ corresponds to a distribution
function $n_{\bf p}(t,{\bf x})$ which is one inside the Fermi line and zero
outside the line.

We now derive the commutation
relation between the $u_n$s with the following prescription.
If we define an operator $F$ (and similarly $G$) as
\begin{equation}
  F = \int\! {d{\bf x}\,d{\bf p} \over (2\pi)^2}\, F({\bf x},{\bf p}) n_{\bf p}
  ({\bf x}),
\end{equation}
where $n_{\bf p}({\bf x})$ is the quasiparticle distribution function,
then we need to impose the condition on the commutation relation so that 
\begin{equation}\label{FG-commut}
  [F,\, G] = -i \!\int\! {d{\bf x}\,d{\bf p} \over (2\pi)^2}\,
  \{F,\, G\} ({\bf x}, {\bf p}) n_{\bf p}({\bf x}),
\end{equation}
where the $\{F,\,G\}$ is the classical Poisson bracket between $F$ and $G$,
\begin{equation}
  \{F,\, G\} = {\partial F \over \partial p_i} {\partial G \over \partial x_i}
  - {\partial G \over \partial p_i} {\partial F \over \partial x_i}
  - b\epsilon^{ij} {\partial F \over \partial p_i}{\partial G\over \partial p_j}
  ,
\end{equation}
where we have allowed the composite fermions to be in an external
magnetic field $b$. For Jain's sequences $b=\pm B/(2N+1)$.
Restricting $n_{\bf p}$ to be of the form of the step function (1
inside the Fermi line and 0 outside), $F$, $G$, and the right-hand side of Eq.~(\ref{FG-commut}) become
functionals of the shape of the Fermi surface, and
one can easily derive
the commutator of the small perturbations $u$:
%For the purposes of this paper, it is sufficient to find the
%commutator of $u$'s to order $O(u^0)$.  For that, we expand $F$ and
%$G$ to linear order in $u$,
%$$
%  F = F_0 + {p_F \over 4\pi^2}\! \int\!d{\bf x}\,d\theta\,
%  F({\bf x}, \theta) u({\bf x},\theta) + O(u^2)
%$$
%where $F_0=F(un=0)$ is a $c$-number (and a similar equation for $G$), and
%compute the right hand side of Eq.~(\ref{FG-commut}) to order $O(u^0)$.  For
%the latter, we integrate by part with respect to ${\bf p}$, using 
%$\partial n_{\bf p}/\partial{\bf p}=-{\bf n}(\theta)\delta(p-p_F)$.
%We find, at the end,
\begin{multline}
  [u({\bf x},\theta),\, u({\bf x}',\theta')] =
  {i(2\pi)^2 \over p_F} \biggl(
    -n_i(\theta){\partial \over {\partial x_i}}
    + {b\over p_F}{\partial \over \partial\theta}
      \biggr)\\  [\delta({\bf x}-{\bf x}')\delta(\theta-\theta')]
   +O(u),
\end{multline}
where ${\bf n}(\theta)=(\cos\theta,\,\sin\theta)$.  In terms of $u_n$, the
formula reads   
%$$
%  [u_m({\bf x}),\, u_n({\bf x}')] =
%  - {2\pi  \over p_F^2}\biggl[b m \delta_{m+n,0}
%    + {i\over 2} p_F \delta_{m+n,1} (\partial_x - i\partial_y)
%    + {i\over 2} p_F \delta_{m+n,-1} (\partial_x + i\partial_y)
%  \biggr]  \delta({\bf x}-{\bf x}')
%$$
\begin{multline}\label{commutator}
[u_m({\bf q}),\, u_n({\bf q}')] =
   {\pi  \over p_F}\biggl[{2b m\over p_F} \delta_{m+n,0}
    +  \delta_{m+n,1} q_+\\
    + \delta_{m+n,-1} q_-
  \biggr]  (2\pi)^2\delta({\bf q}+{\bf q}') + O(u),
\end{multline}
%  \hspace{-7.5cm} + O(u^2).
where $q_\pm=q_x\pm iq_y$.  This commutation relation has been
previously derived in Ref.~\cite{Haldane:1994} by extending Tomonaga's
bosonization method to higher dimensions.
Note that the algebra depends only on the
size of the Fermi surface $p_F$ but not on any dynamic properties
(Fermi velocity, Landau's parameters etc.).  

{\em Gauging the Fermi surface.}---The composite fermion is coupled to
a dynamical gauge field.  
%If the electron-electron interaction is
%short-ranged then the gauge field has a local action, and composite
%fermions form a non-Fermi liquid.  On the other hand, when the
%electron-electron interaction is longer ranged than the Coulomb
%interaction, the composite fermions form a Fermi liquid.  For Coulomb
%interaction, one has a marginal Fermi liquid with logarithmically
%running parameters~\cite{Halperin:1992mh}.  The last two cases will be
%analyzed explicitly in what follows, but will argue that some of our
%conclusions survive even in the non-Fermi liquid case.
A Fermi surface coupled to a gauge field is a long-standing theoretical
problem, and the bosonized language allows us to partly address it.

In the bosonic description, the temporal component of the gauge field
$a_0$ is coupled to $u_0$, and the spatial components are coupled to
$u_{\pm1}$.
In the Dirac composite fermion theory, the leading term in action for $a_\mu$
is the Maxwell term.
If the dynamical gauge field is at infinitely strong
coupling, then
%the coupling to an external field simply imposes
the constraints $u_0=u_{\pm1}=0$ arise
as the result of the equations of motion $\delta S/\delta a_\mu=0$.
The assumption of strong gauge coupling
should become better and better in the limit $N\to\infty$.  This is
due to two reasons.  First, the coupling of the composite fermions
to the gauge field is set at the Fermi energy $\epsilon_F$ and
momentum $p_F$, while the scales of interest for our problem are
$\epsilon_F/N$ and $p_F/N$.  This gauge coupling is relevant for
contact and marginal for Coulomb interactions.  Second, at these low
energies the Fermi surface is effectively $O(N)$ fermionic species
(corresponding to $O(N)$ patches on the Fermi surface in the
renormalization group
treatment~\cite{Polchinski:1992ed,Shankar:1993pf}), boosting the 't
Hooft coupling by an additional factor of $N$.
(The argument is more complicated in the case of the HLR theory with a
Chern-Simons term in the action for $a_\mu$, but the conclusion is the
same).

{\em Hamiltonian and equation of motion}.---Assuming the composite
fermions form a Fermi liquid with Landau's parameters $F_n$, the
Hamiltonian of the system is
\begin{equation}\label{Hamiltonian}
  H = {v_F p_F \over 4\pi}\!\int\!d{\bf x}\!\sum_{n=-\infty}^\infty\!
  (1+F_n)u_n({\bf x}) u_{-n}({\bf x}),
\end{equation}
where $F_n$ are the Landau parameters.  In the case of a marginal Fermi liquid, we may understand by $F_n$ the Landau parameters evaluated at the scale of the
energy gap.  The Hamiltonian~(\ref{Hamiltonian}) and the commutation
relations~(\ref{commutator}) form our theory of the neutral
excitations in the fractional quantum Hall fluid.  This theory involves
an infinite number of fields $u_n$, reminiscent of higher-spin
relativistic field theories~\cite{Fradkin:1987ks,Fradkin:1986qy}.

Let us first consider a zero wave number.  Then according to
Eq.~(\ref{commutator}) the operators $u$ can be divided into pairs of
creation and annihilation operators $(u_{-2},u_2)$, $(u_{-3},u_3)$,
etc., with $u_n$ for $n>0$ being the annihilation and with $n<0$,
creation operators.  The frequency of the oscillators are
\begin{equation}\label{omega0}
\omega_n^{(0)} = n (1+F_n)  \omega_c, \quad \omega_c = {b \over m_*}\,.
%\equiv {b v_F\over p_F}, \quad n=2,3, \ldots
\end{equation}
The index $n$ can be interpreted as the spin of the excitation.  For
example, the contribution of spin-$n$ mode to the spectral density of
the density operator is expected to be $q^{2n}$ at small $n$, so the leading
contribution to the spectral weight comes from the $n=2$ mode.   
The ordering in energy of the modes depends on $F_n$; in the simplest scenario
$n=2$ is the lowest mode.  Since $\omega_c\sim N^{-1}$, and the cutoff
of our theory is $O(N^0)$, one should expect $O(N)$ of these modes (provided
that $F_n$ does not increase as a power of $n$).

If one puts $F_n=0$ in Eq.~(\ref{omega0}), one would find
$\omega_n^{(0)}=n\omega_c$.  This can be interpreted as the energy of
creating a pair of a quasiparticle and a quasihole, separated by $n$
Landau-level steps.  Note that the na\"ive lowest mode with $n=1$ disappears
due to the coupling to the dynamical gauge field~\cite{Kim:1995}.  As
far as we know, Eq.~(\ref{omega0}) does not have a simple
interpretation when the Landau parameters are nonzero.

To find the dispersion relation at finite wave number $q$, one needs to
solve the linearized equation of motion, which can be obtained by
taking the commutator with the Hamiltonian~(\ref{Hamiltonian}).
In momentum space, choosing ${\bf q}$ to point along the $x$ axis, the
equation is
\begin{multline}\label{EOMF}
  [\omega - n(1+F_n)\omega_c] u_n =  {v_Fq\over 2}[ (1+F_{n-1}) u_{n-1}\\
    + (1+F_{n+1}) u_{n+1}]
\end{multline}
for $n\ge2$ and $n\le-2$ and where by construction $u_{\pm 1}=0$.
The task of finding the spectrum of excitations thus reduces to finding
the eigenvalues of a certain tridiagonal matrix.
Using Eq.~(\ref{omega0}), this equation can be rewritten as
\begin{equation}\label{EOM}
  (\omega-\omega_n^{(0)})u_n = {2N+1\over2} q\ell_B \biggl[
    {\omega_{n-1}^{(0)} \over n-1}u_{n-1}
   + {\omega_{n+1}^{(0)} \over n+1}u_{n+1} \biggr].
\end{equation}
 Remarkably,
Eq.~(\ref{EOM}) determines completely the dispersion curves from their starting points at $q=0$.  Thus we speculate that Eq.~(\ref{EOM}) is valid even when
the $\nu=1/2$ state is a non-Fermi liquid.
For small
$q$ the equation can be solved perturbatively over $q$.  For example, for the $n=2$ mode we find
\begin{equation}\label{curvature}
{\omega_2(q) \over \omega_2^{(0)}} = 1
- {(2N+1)^2 \over 24 \bigl(1-\omega_2^{(0)}/\omega_3^{(0)}\bigr)} (q\ell_B)^2
%  {\omega_3^{(0)} \over \omega_3^{(0)}-\omega_2^{(0)}}
  + O(q^4).
\end{equation}
If the spin-2 mode is the lightest one, then its dispersion curve
bends down when we go to finite $q$.  Equation~(\ref{curvature}) relates 
the curvature at $q=0$ of the lowest mode and the ratio of
the energies of the spin-3 and spin-2
modes, and is one prediction of the theory.
%Note that nontrivial momentum dependence appears for
%$q\ell_B\sim O(N^{-1})\ll1$, within the regime of validity of our theory.
%For $q\ell_B\sim N^{-1}$, the perturbation expansion
%over $q$ cannot be used, but knowing $F_n$ one can always determine
%numerically the spectrum of excitations.

It is intriguing that Ref.~\cite{Hirjibehedin:2005} found two modes at
$\nu=1/3$.  While it is tempting to identified them with spin-2 and
spin-3 excitations, it is unclear if such an identification can be made
at such a low value of $N$, $N=1$.

\begin{figure}
  \centering
  \includegraphics[width=24em]{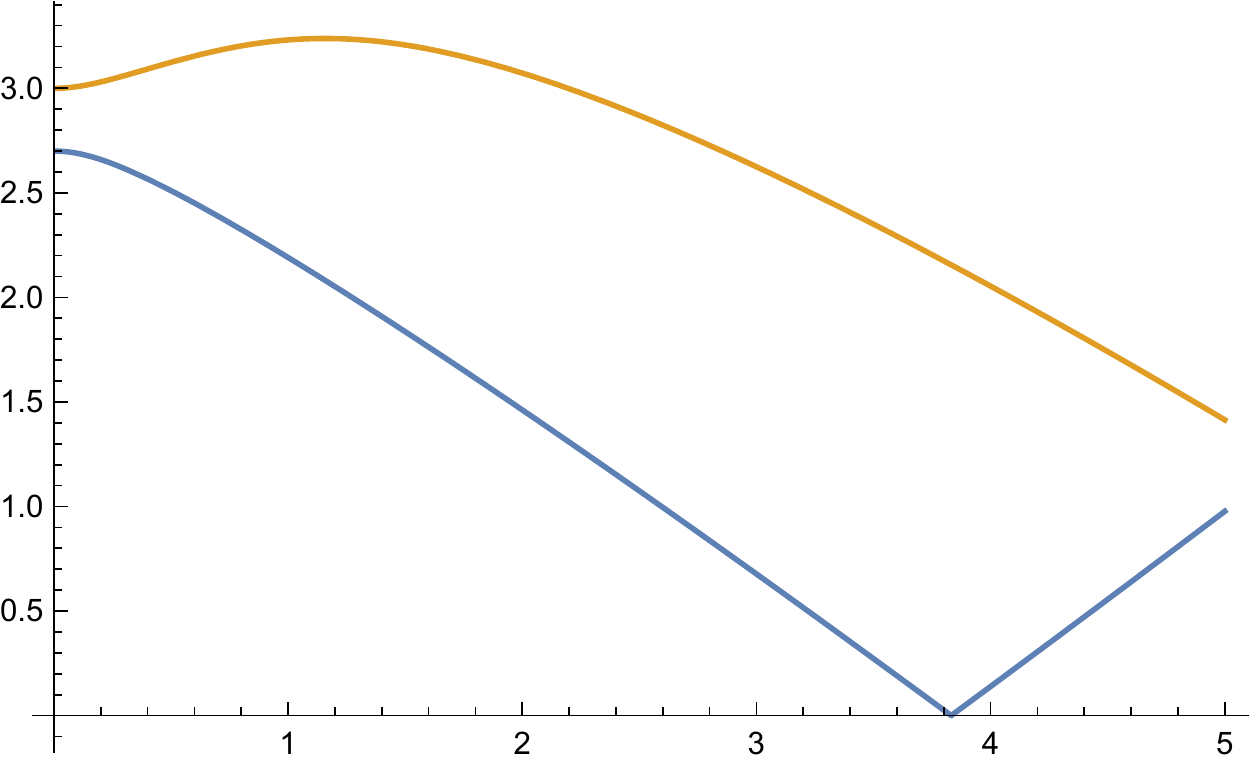}
  \caption{The dispersion curves for the lowest two modes for $F_2=0.35$, $F_n=0$ with $n\ge3$.  The horizontal axis is $(2N+1)q\ell_B$ and the vertical axis is the energy in units of $\omega_c$.  The cusp at zero energy is an artifact of the infinite $N$ limit.}
  \label{fig:dispersion}
\end{figure}

{\em The magnetoroton minima.}---For $Nq\ell_B\sim1$ one has to solve
the full system of equations, Eq.~(\ref{EOMF}) or (\ref{EOM}), to find
the dispersion curves.  In Fig.~\ref{fig:dispersion}, we plot a
typical result.  We note that the energy of the lowest mode goes to
zero at a finite momentum.  We now show analytically that this always
happens at an infinitely strong gauge coupling. 
We need to solve Eq.~(\ref{EOM}) with $\omega=0$ and the boundary
conditions $u_1=0$ and $u_n\to0$ when $n\to\infty$.  The solution to this
recursion relation, which satisfies the boundary condition $u_n\to0$
when $n\to\infty$, is
\begin{equation}\label{zeromode}
  u_n = {(-1)^n \over 1+F_n} J_n\Bigl({p_Fq \over b}\Bigr) .
\end{equation}
The boundary condition $u_1=0$ requires $J_1(p_Fq/b)=0$.
The latter occurs at $q=z_ib/p_F$, where $z_i$ are the zeros of the
Bessel function $J_1$.  One can write this as
\begin{equation}\label{locations}
  q\ell_B = z_i \frac b{p_F^2} = z_i \frac bB = \frac{z_i}{2N+1}
\end{equation}
for the filling fractions $\nu=N/(2N+1)$ and $\nu=(N+1)/(2N+1)$.

The fact that the energy
of an excitation is exactly zero is an artifact of the
strong gauge coupling approximation, which we have argued to occur at
infinite $N$; when the hard constraints on $u_0=u_{\pm1}=0$ are
relaxed, these zeros of the dispersion relation should become
minima.  The values of the energy at the minima are
smaller by a power of $N$ compared to the energy scale of the excitations at $q=0$ ($\omega_n^{(0)}$) but are nevertheless nonzero~\footnote{This qualitative
  feature has been seen previously in an improved RPA calculation within the
  HLR theory~\cite{Simon:1993}}.
%The minima should be parametrically deep in the limit of
%large $N$: the value of the energy at the minima should be smaller by
%a power of $N$ compared to the energy scale of the excitations at
%$q=0$ ($\omega_n^{(0)}$), but nonzero.
This is confirmed in a more detailed treatment of the composite fermions,
taking into account the density-density Coulomb interaction~\cite{GNRS2}. 
On the other hand, the strict
$N=\infty$ limit of infinitely strong gauge coupling allows us to
determine analytically the locations of the minima of the dispersion curves.
Here we find a surprising result that the positions of the minima on the momentum axis do
not at all depend on the parameters appearing in the
Hamiltonian~\footnote{In Ref.~\cite{Halperin:1992mh} it was noticed
  that the minima in the longitudinal conductivity $\sigma_{xx}$ occur when
  $(2N+1)q\ell_B$ is near the zeros of the Bessel function $J_1$.}.

We now show that 
the robustness of the locations of the magnetoroton minima is due to
them being determined by the commutator
algebra~(\ref{commutator}) but not by the Hamiltonian.  In fact, at the
values of $q$ set by Eq.~(\ref{locations}), there exists a pair of
operators $\hat O$ and $\hat O^\dagger$, which commutes with all
$u_n$ (and consequently with the Hamiltonian) to leading order in $u$:
\begin{equation}
  \hat O = \sum_{n=2}^\infty (-1)^n J_n \biggl( {p_F\over b} q \biggr) u_n .
\end{equation}
In other words, if one defines the commutator matrix $C_{mn}$ as
\begin{equation}
  [u_m({\bf q}), \, u_{-n}({\bf q'})] = C_{mn} (2\pi)^2\delta({\bf q}+{\bf q}')
\end{equation}
for $m,n>0$, where
\begin{equation}
 C_{mn} = {2\pi b\over p_F^2} \begin{pmatrix} 2 & z & 0 & 0 & \ldots \\
  z & 3 & z & 0 & \ldots \\
  0 & z & 4 & z & \ldots \\
  0 & 0 & z & 5 & \ldots \\
  \ldots & \ldots & \ldots & \ldots & \ldots
  \end{pmatrix}, ~~ z = {2N+1\over 2} q\ell_B,
\end{equation}
then at the momenta~(\ref{locations}) the matrix $C$ has a zero
eigenvalue.  Across these momenta, the role of creation and annihilation
operators is exchanged for one pair of operators.  It is not difficult
to show that any Hamiltonian quadratic in $u$'s needs to have a zero
eigenvalue when such an exchange occurs.

The positions of the magnetoroton minima~(\ref{locations}) and their
complete independence of the details of the Hamiltonian are the
central result of this Letter.  In the past, model calculations have
shown that the positions of the magnetoroton minima depend very
weakly on the interactions (see, e.g., Ref.~\cite{Balram:2015pye}),
but the fundamental reason behind this fact was not understood.

It is worth remembering, however, that our derivation requires
$q\ell_B\ll 1$, which means that $z_i$ in Eq.~(\ref{locations}) should
be one of the first $o(N)$ roots of $J_1$.
%Our calculation does not
%give, e.g., the total number of magneto-roton minima.
However,
the values found in Eq.~(\ref{locations}) seem to fit the existing
data quite well even for relatively large $q\ell_B$.
%Ref.~\cite{Kukushkin:2009} gives one minimum for $\nu=2/5$, two for
%$\nu=3/7$, and three for $\nu=4/9$ (it was suggested that the
%experiment missed the last minimum with highest wave number at each
%filling fraction).
Limiting ourselves to the range explored in
Ref.~\cite{Kukushkin:2009}, $q\ell_B\lesssim1.2$,
our prediction for the locations of the magnetoroton
minima is summarized in the following table
(experimental values extracted from Ref.~\cite{Kukushkin:2009} in parentheses):
\smallskip
%\begin{itemize}
%\item $\nu=2/5$: $q_1\ell_B=0.77$ (exp: 0.86);
%\item $\nu=3/7$: $q_1\ell_B=0.55$ (0.52), $q_2\ell_B=1.00$ (1.06);
%\item $\nu=4/9$:
%  $q_1\ell_B=0.43$ (0.40), $q_2\ell_B=0.78$ (0.85), $q_3\ell_B=1.13$ (1.25).
%\end{itemize}

%\begin{table}
\begin{tabular}{|c|c|c|c|}
  \hline
  & $n=1$ & $n=2$ & $n=3$\\
  \hline
  $\nu=2/5$ & 0.77 (0.86) & & \\
  \hline
  $\nu=3/7$ & 0.55 (0.52) & 1.00 (1.06) & \\
  \hline
  $~\nu=4/9~$ & $~$0.43 (0.40)$~$ & $~$0.78 (0.85)$~$ & $~$1.13 (1.25)$~$ \\
  \hline
\end{tabular}
%\end{table}
\smallskip

\noindent
All these values
are surprisingly close (within 15\% or less) to existing
experimental~\cite{Kukushkin:2009} and
numerical~\cite{Scarola:2000} results, despite the smallness of $N$
and the large values of the $q\ell_B$ under discussion.  Even for
$N=1$, the calculated position of the magnetoroton $q\ell_B=1.28$ is
in good agreement with the original estimate of Ref.~\cite{Girvin:1986zz}.
We interpret the agreement as confirming the validity of the
interpretation of the low-lying neutral excitations as shape fluctuations
of the Fermi surface.

Since the locations of the magnetoroton minima depend only on the
commutator algebra, which originates from the kinematics of the Fermi
surface rather than from the Hamiltonian, we expect the minima would
survive even in the non-Fermi-liquid regime of short-ranged
electron-electron interactions.

In summary, the universal momenta at the magnetoroton
minima~(\ref{locations}), along with the existence of multiple
branches of neutral excitations, each with a distinct value of the
spin at $q=0$, are the main predictions of this Letter.  These
predictions should be valid in any system described by a Fermi surface
coupled to a dynamical gauge field in a small background magnetic
field.

The authors thank Jainendra Jain, David Mross, Subir Sachdev, and
Paul Wiegmann for discussions.
This work is supported, in part, by U.S. DOE Grant
No. DE-FG02-13ER41958.  D.~T.~S. is supported, in part, by a Simons
Investigator Grant from the Simons Foundation.  S.~G. is supported, in
part, by European Research Council under the European Unions Seventh
Framework Programme (FP7/2007-2013), ERC Grant Agreement No.\ STG 279943,
``Strongly Coupled Systems.''  Additional support was provided by the
Chicago MRSEC, which is funded by NSF through Grant No.\ DMR-1420709, and
by the ARO MURI grant No.\ 63834-PH-MUR.

\bibliography{higherspin-final}

\end{document}